%% file: bulk_apj.tex
\documentclass[preprint,useAMS,usegraphicx]{aastex}
\usepackage{natbib}
\usepackage[dvips]{color}
\input{definitions}

\def\ace{ASCE}
\shorttitle{The Cosmological Bulk Flow}
\shortauthors{Nusser \& Davis}
\begin{document}
\title{The cosmological  bulk flow: consistency with  $\Lambda$CDM and $z\approx 0$ 
constraints on $\sigma_8$ and $\gamma$}
   \author{Adi Nusser\altaffilmark{1}}
\affil{Physics Department and the Asher Space Science Institute-Technion, Haifa 32000, Israel}
\author{Marc Davis\altaffilmark{2}}
\affil{Departments of Astronomy \& Physics, University of California, Berkeley, CA. 94720}
\altaffiltext{1}{E-mail: adi@physics.technion.ac.il}
\altaffiltext{2}{E-mail: mdavis@berkeley.edu}

\begin{abstract}
We derive estimates for the  cosmological bulk flow from  the SFI++  Tully-Fisher (TF) catalog.   For a  sphere of radius $40 \hmpc$ centered on the MW, we derive 
a bulk flow of $333 \pm 38\kms $ towards   Galactic $ (l,b)=(276^\circ,14^\circ)$ within a $3^\circ$ $1\sigma$ error. 
Within a  $ 100\hmpc$ we get $ 257\pm 44\kms$  towards $(l,b)=(279^\circ, 10^\circ)$ within a $6^\circ$ error. 
These directions are at a $40^\circ$  with the Supergalactic plane, close 
 to the apex of the motion of the Local Group  of galaxies  after the   Virgocentric infall correction.
Our findings are  consistent with the $\Lambda$CDM model with 
the latest WMAP best fit  cosmological parameters.  But the bulk flow allows 
independent constraints.
For  WMAP inferred  Hubble  parameter $h=0.71$ and  baryonic mean density parameter $\Omega_b=0.0449$, the  
constraint from the bulk flow on the matter  density  $\Omega_m$, the  normalization of the density fluctuations, $\sigma_8$, and the growth index, $\gamma$, can be expressed as   
$\sigma_8\Omega_m^{\gamma-0.55}(\Omega_m/0.266)^{0.28}=0.86\pm 0.11$ (for $\Omega_m\approx 0.266$). Fixing 
 $\sigma_8=0.8$ and $\Omega_m=0.266$ as favored by WMAP, we get $\gamma=0.495\pm 0.096$.
The constraint derived here rules out popular  DGP models at more than the 99\% confidence level.
Our results are  based on a method termed \ace\ (All Space Constrained Estimate)  
 which reconstructs the bulk flow from an all space three dimensional peculiar velocity field 
constrained to match the TF measurements. 
At large distances \ace\ generates a robust bulk flow from the SFI++ that is insensitive 
to the assumed prior. For comparison, a standard straightforward  maximum likelihood estimate  
leads to very similar results.

\end{abstract}

\keywords{Cosmology: large-scale structure of the Universe, dark matter, cosmological parameters}

\section{Introduction}
Cosmological bulk flows are  the peculiar velocities of whole spherical regions   around us. 
 Bulk flows are usually considered for sufficiently large spheres where linear expressions for 
the velocity and density power spectra are valid. This greatly facilitates the calculation of expected 
bulk flows in cosmological models, in contrast to analyzing the full field field which may 
involve non-linear effects on small scales \citep{feldwh10, abate09, zarou09, freu99}.
In linear theory,  the bulk flow of a sphere is solely determined by the gravitational pull
of only the dipole component of the external mass distribution. 
Bulk flows  are, therefore,  an unmistakable indicator of distant  large mass concentrations should they exist. 
 The exact expression  of the  bulk flow, $\vB(r)$, of a sphere of radius $r$ is,
 \begin{equation}
\label{eq:bdef}
\vB(r)=\frac{3}{4\pi r^3}\int_{x<r}  \vv({\vx})  \dd^3 x  \; .
\end{equation}
where $\vv(\vx)$  is the 3D peculiar velocity field  as a function of the comoving coordinate $\vx$.
Beneath this   innocuous expression lie  a multitude of nuisances.
An unbiased estimate of $\vB$ requires knowledge of $\vv$ sampled uniformly overall 
the volume. However,  observational probes of peculiar velocities  measurements  are  available only 
for a few thousand galaxies with a patchy coverage of the local Universe.  Further, peculiar velocity probes such as the TF relation
allow us to constrain only the radial component of the peculiar velocities of galaxies.

Recently compiled data on peculiar velocities has triggered renewed interest in the analysis of large scale flows, including the bulk flow
\citep{DN10, l10, Erdo06}.
\cite{feldwh10} report an unusually large bulk flow of $416\pm 78\kms$  in a sphere of $ 100\hmpc$ which is 
at odds with the $\Lambda$CDM model with the best fit parameters of 
the Seven-Year Wilkinson Microwave Anisotropy Probe  (WMAP7)   \cite[e.g.][]{jaro10,wmap7}.
Here we provide an alternative estimate  of the bulk using a single data set of TF measurements of galaxies, trimmed
at faint magnitudes to ensure the linearity of the TF relation. The estimate is based on a method which we term 
\ace\ for All Space Constrained Estimate. The method  computes $\vB(r)$ using  (\ref{eq:bdef}) from a three dimensional $\vv(\vx)$ defined 
everywhere in a large region of space and constrained to match the TF data.
For the analysis below, we use the SFI++  survey of spiral
galaxies with I-band Tully-Fisher distances,  \citep{mas06, spring07}, 
which  builds on the original Spiral Field
I-band Survey \citep{ gio94, gio95, hg99} and
Spiral Cluster I-band Survey \citep{gio97a, gio97b}.  We use
the published SFI++ magnitudes and velocity widths, and derive our own
peculiar velocities, rather than taking the published distances as given.
We shall use the {\it inverse} of the Tully-Fisher (ITF)
relationship. 
The main advantages of ITF methods is that  samples selected by magnitude, as most are, will be minimally plagued by   Malmquist bias effects when analyzed in the inverse
direction  \citep{schechter, a82}.
  We assume that the circular
velocity parameter, $\eta \equiv {\rm log} (line\; width)$, of a galaxy is, up to a random
scatter, related to its absolute magnitude, $M$, by means of a
linear {\it  inverse} Tully-Fisher (ITF) relation, i.e.,
\begin{equation}
\label{eq:ITF}
\eta=s M + \eta_0 .  
\end{equation}
 The preparation of the data is 
done  following  \cite{DN10}. 
We
include all field, group, and cluster galaxies. Galaxies in groups and clusters are
treated as individual objects, though the redshifts for template cluster
galaxies are replaced by the systematic redshift of the cluster.  Galaxies 
fainter the an estimated magnitude of -20 were removed from the sample 
as those showed significant deviations from a linear TF relation. 
In order to get a cleaner TF
sample we select only objects with inclination $i > 45^\circ$  to ease
problems with inclination corrections.  All this leaves us with a sample of 2859 galaxies
with redshifts less than $100\hmpc$.
The effective depth of the sample defined as the error weighted mean redshift of galaxies is $\sim 40\hmpc$.

 We will refer with $\Lambda$CDM7 to the $\Lambda$CDM cosmological model with the WMAP7 best fit parameters 
 \citep{wmap7} for a flat 
Universe, i.e. the 
total mass density parameter $\Omega_m=0.266$, baryonic density parameter $\Omega_b=0.0449$, 
a Hubble constant $h=0.71$ in units of  $100\kms {\rm Mpc}^{-1}$,  a scalar spectral index $n_s=0.963$, and 
$\sigma_8=0.8$ for the rms of linear density fluctuations in spheres of $8\hmpc$. 
Throughout the paper,  variants of $\Lambda$CDM7 with different $\Omega_m$ and $\sigma_8$ will 
be considered. All other parameters will be fixed at their WMAP7 values. 

The outline of the paper is as follows.  Details of  the \ace\ method are described in \S\ref{sec:ace}, while the more
standard Maximum Likelihood Estimate (MLE)  outlined in \S\ref{sec:mle}.
Tests of the methods using mock catalogues designed to match the SFI++ catalogue are presented
in \S\ref{sec:mle}. Results for the bulk flows from the SFI++ data are given in \S\ref{sec:res} with the subsection 
\S\ref{sec:comp} providing a comparison with the $\Lambda$CDM models. 
Finally, \S\ref{sec:disc} discusses  the results and some of their cosmological  implications.

\section{The All Space Constrained  Estimate (\ace)}
\label{sec:ace}
Observations of distance (peculiar velocity) indicators, such as the SFI++ TF survey, are available for 
only a small fraction of galaxies in the local Universe (out to $\sim 100\hmpc$).  The absence of uniformly distributed data 
prevents a direct application of equation (\ref{eq:bdef}). To circumvent this 
problem, the \ace\ method   effectively uses (\ref{eq:bdef}) to reconstruct the 
bulk flow from a  3D field $\vv({\vx})$ which satisfies  two conditions:  a) at sufficiently large distances from  the observed galaxies in the TF data, 
it has a power spectrum that is dictated by a  cosmological model such as the $\Lambda$CDM, and  b) it has 
 radial peculiar velocities at the positions of the observed galaxies, which  are consistent 
 with the TF measurements.
The approach is similar to that of constrained realization from noisy data \citep{hr91,zh95}, but it is more general and easier to 
implement.
Assume that the TF catalogue contains $i=1\dots N_{\rm g}$ galaxies with measured redshifts  (in $\kms$), $cz_i$,
apparent magnitudes, $m_i$, and line width parameters, $\eta_i$. 
We write the absolute magnitude of a galaxy, 
\begin{equation}
M_i = M_{0i} + P_i\; ,
\end{equation}
where
\begin{equation}
M_{0i} = m_i + 5{\rm log}(cz_i)-15 \; 
\end{equation}
and 
\begin{equation}
\label{eq:utp}
P_i = -5 {\rm log}(1- u_i/cz_i)
\end{equation} 
with $u_i$ the radial peculiar velocity of the galaxy. Both $cz_i$ and $u_i$ are defined  in the frame of the cosmic microwave background radiation (CMB).  
Assume that  an estimate of the 
underlying cosmological velocity field, $\vv(\vx)$, can be written as a linear combination of 
\begin{equation}
\label{eq:linc}
\vv(\vx)=\sum_\alpha^{N_{\rm a}} a^\alpha \vv^\alpha(\vx)\; , 
\end{equation}
where the $N$ fields, $\vv^\alpha(\vx)$ ($\alpha=1\cdots N_{\rm a}$), are   gaussian random velocity fields
generated using a cosmologically viable power spectrum.  In practice these basis velocity  fields will 
be extracted from a linear cosmological velocity field generated 
in a very large box using the power spectrum of the $\Lambda$CDM model. We then compute 
 $u^\alpha_i$, the radial component of the  $\vv^\alpha$ at the redshift space positions 
of the observed galaxies, and 
 define the $P-$basis functions, $P^{\alpha}_i=-5{\rm log}(1-u^\alpha_i/cz_i)$ for the observed galaxies only.
The model $P$ is then written as
\begin{equation}
P^M_i=\sum_\alpha a^\alpha P^\alpha_i\; . 
\end{equation}
The best fit mode coefficients $a^\alpha$, the slope, $s$, and  the zero point
$\eta_0$, are found by  minimizing the $\chi^2$ statistic
\begin{equation}
\label{eq:chi}
\chi^2=\frac{1}{\ssigint}\sum_{i=1}^{N_{\rm g}}
 {\left(s M_{0i}+ s P^M_i+\eta_0-\eta_i\right)^2}+
 \sum_{\alpha=1}^{N_{\rm a}}  (a^\alpha)^2\; ,
\end{equation}
where $\ssigint$ is the $rms$ of the intrinsic scatter in $\eta$ 
about the ITF relation, and $N_{\rm g}$ is the number of galaxies in the sample.
The second term of the sum over the squares of $a^\alpha$ is introduced in order to regularize the solution especially 
in regions of poor data coverage. 
In the appendix we derive this term from a Bayesian  formulation.    
The solution to the equations $\partial \chi^2/\pa a^\alpha=0$, 
 $\partial \chi^2/\pa s=0$ and  $\partial \chi^2/\pa \eta_0=0$ is straightforward.
%
The coefficients $a^\alpha$ will be used in  equation (\ref{eq:linc}) to get  $\vv(\vx)$ everywhere in a region of 
space large enough to contain the data. 
For each field $\vv^\alpha(\vx)$ we compute its corresponding bulk flow, $\vB^\alpha(r)$,  according to equation (\ref{eq:bdef}) and 
write our \ace\   bulk flow as
\begin{equation}
\label{eq:best}
\vB_{_{\rm ASCE}}(r)=\sum_\alpha a^\alpha \vB^\alpha(r)\; .
\end{equation}

\section{The Maximum Likelihood Estimate (MLE)}
\label{sec:mle}
For comparison we will present estimates of the bulk flow obtained with  the standard MLE  \citep{K88}.
This method  approximates the bulk flow of a  sphere of radius $r$ as the vector $\vB_{_{\rm MLE}} $ 
which renders a minimum in 
\begin{equation}
\chi^2=\frac{1}{\ssigint}\sum_{cz_i <r}
{\left(s M_{0i}+ 2.17s \frac{  \vB \cdot \hat {\vr}_i}{ cz_i}+\eta_0-\eta_i\right)^2}\; 
\end{equation}
with respect to the three components of $\vB$. The sum is over galaxies within $r$ and $\hat {\vr}_i$ is  unit vector in the
direction of galaxy $i$. Further, in this expression we have approximated $P=-5{\rm log}(1-\vB \cdot \hat {\vr}_i/cz_i)\approx
 2.17  \vB \cdot \hat {\vr}_i/{ cz_i}$.

\section{Tests}
\label{sec:test}

\begin{figure} 
\centering
\includegraphics[ scale=0.7 ,angle=00]{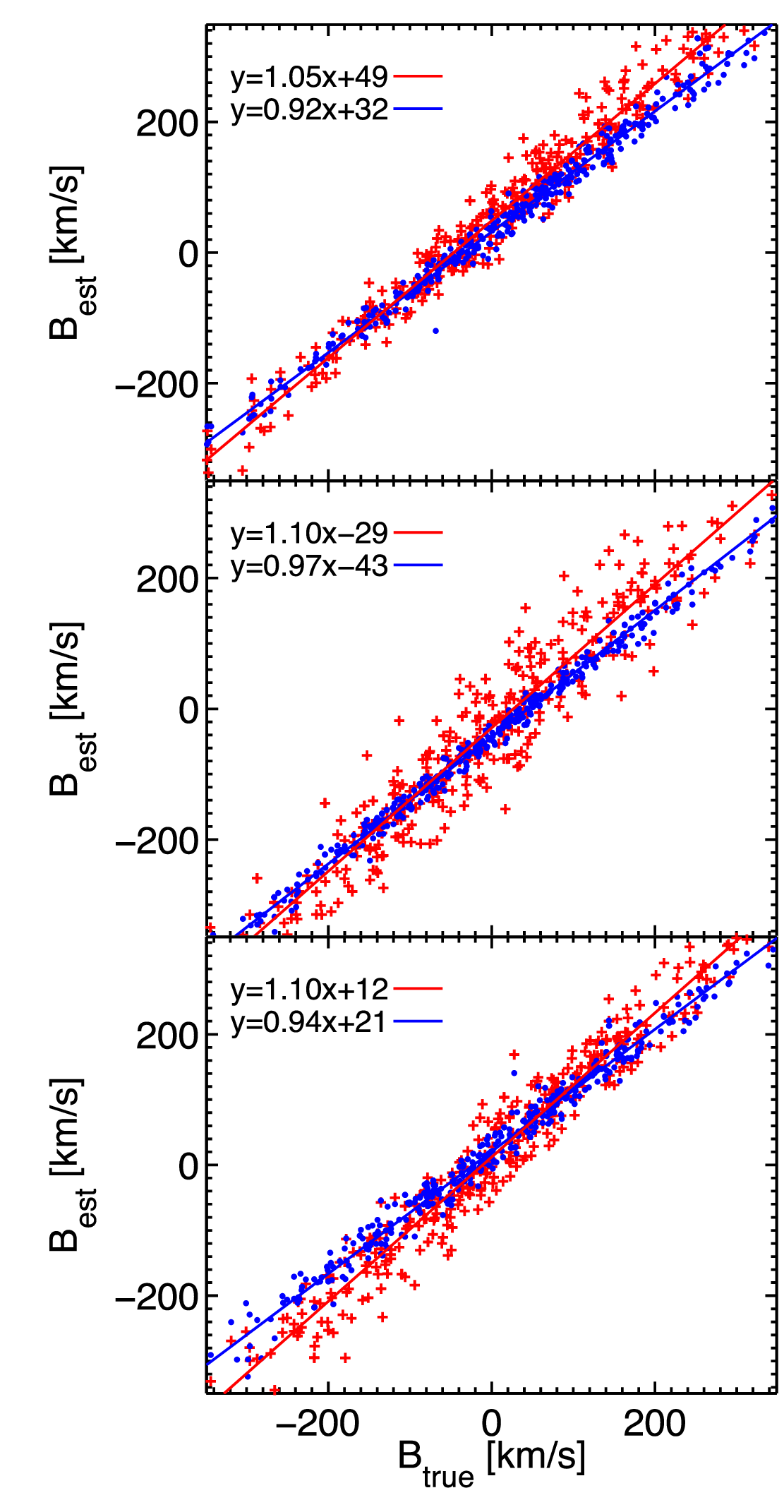}
\caption{Scatter plots of estimated versus true bulk flows in 400 mock catalogues. 
The {\bf top}, {\bf middle} and {\bf bottom} panels correspond to the Supergalactic $x$, $y$ and $z$ components of the bulk flow.
Plotted are bulk flows in  a spherical region of $r=60\hmpc$ centered at the origin. 
Bulk flows from \ace\ and MLE are represented as blue dots and red crosses, respectively.
The overlaid lines in each panel are linear  regressions.  }
\label{fig:scatt}
\end{figure}
    
\begin{figure} 
\centering
\includegraphics[ scale=0.7 ,angle=00]{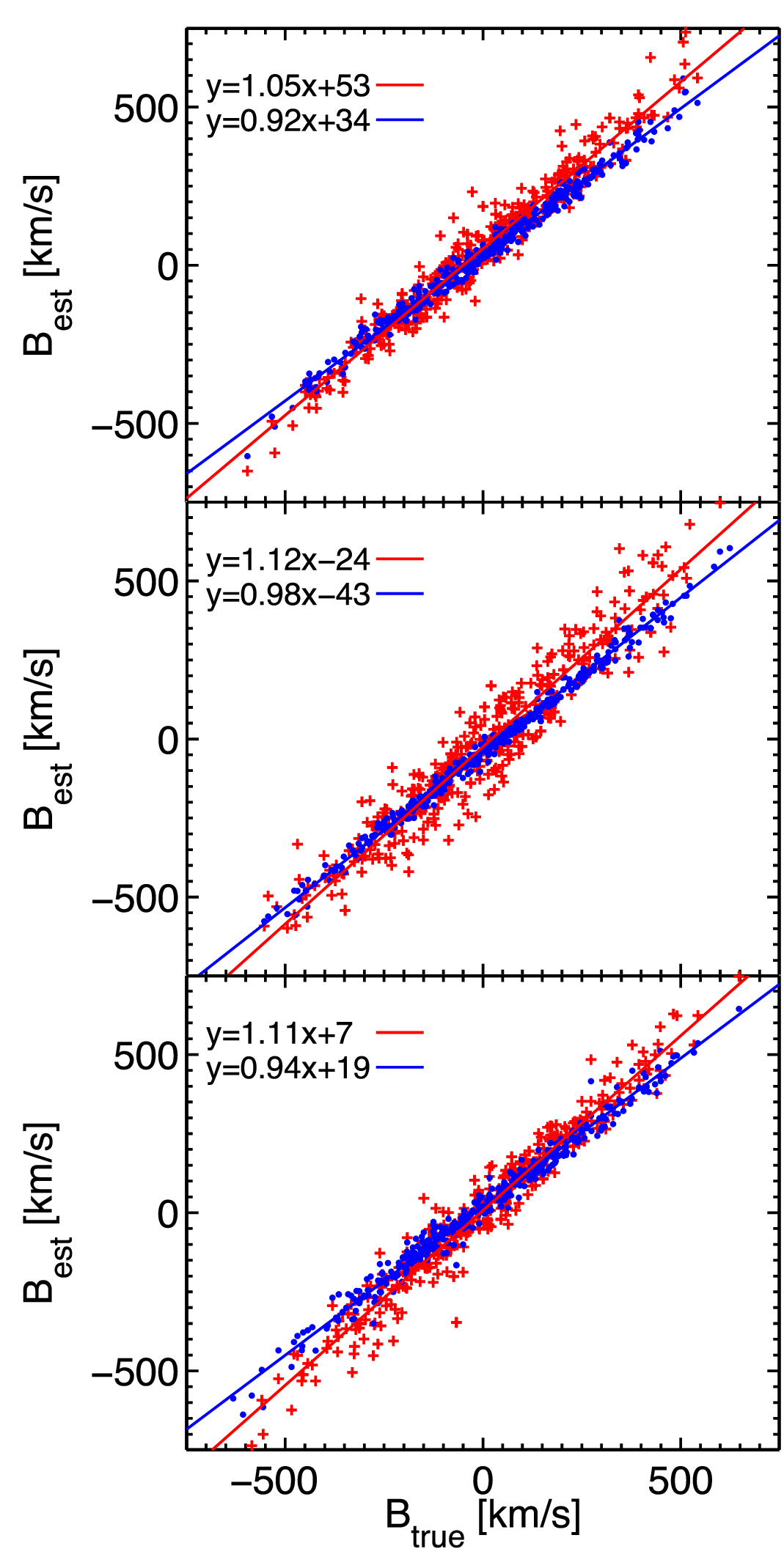}
\caption{The same as the previous figure but where the peculiar velocities  in the mocks have been amplified by a
factor of 1.5.  }
\label{fig:scatt15}
\end{figure}

\begin{figure} 
\centering
\includegraphics[ scale=0.4 ,angle=00]{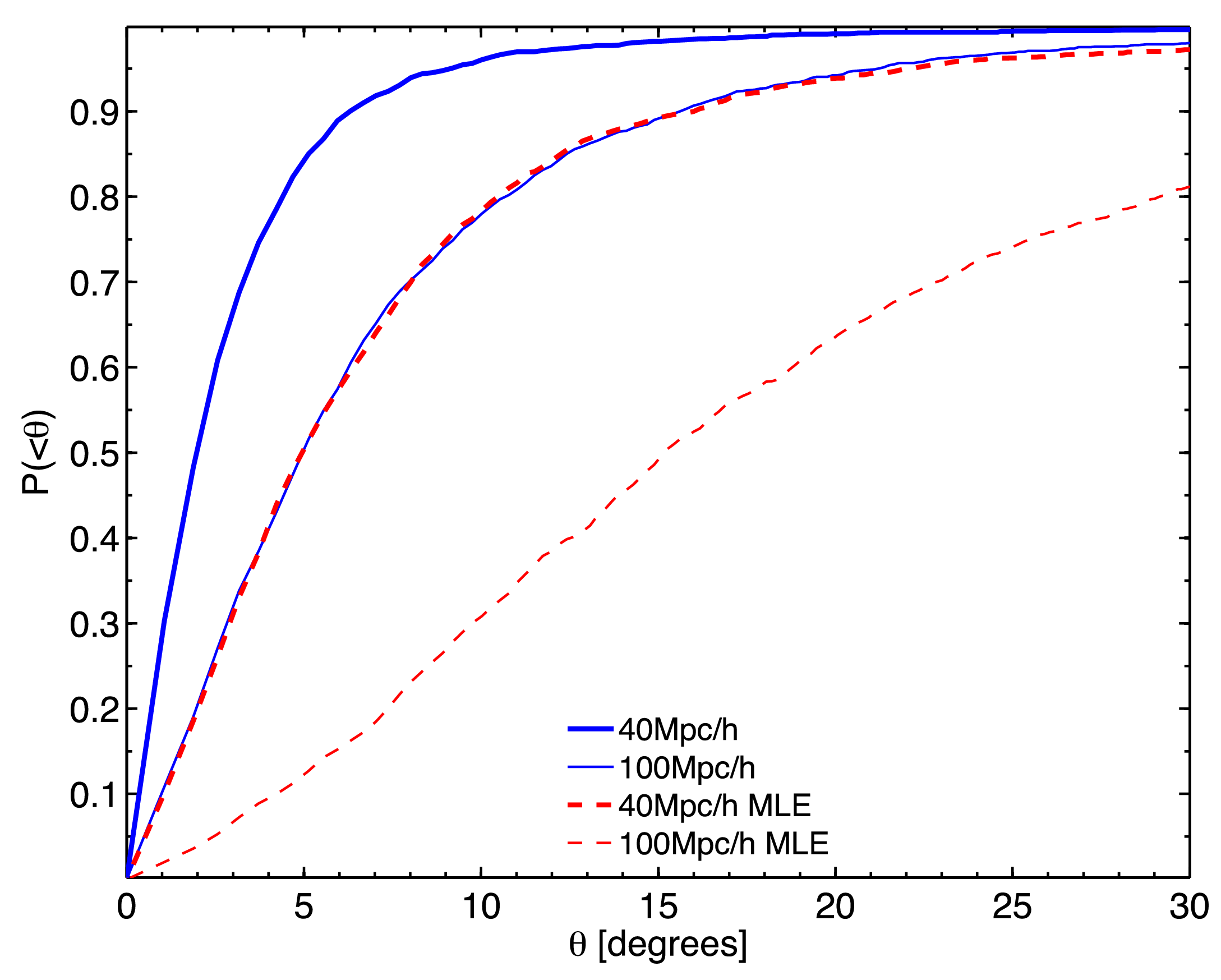} 
\caption{The cumulative fraction of mock catalogues with estimated bulks directed within angle 
smaller than $\theta$ from the direction of the true bulk flow.  Blue solid lines and red dashed lines, respectively, correspond to 
the \ace\  and MLE reconstructions. Thick and thin lines, respectively,  refer to bulk flows of spheres of radii $40 \hmpc$ and  $ 100\hmpc $ centered on the observer. }
\label{fig:costh}
\end{figure}

\begin{figure} 
\centering
\includegraphics[ scale=0.4 ,angle=00]{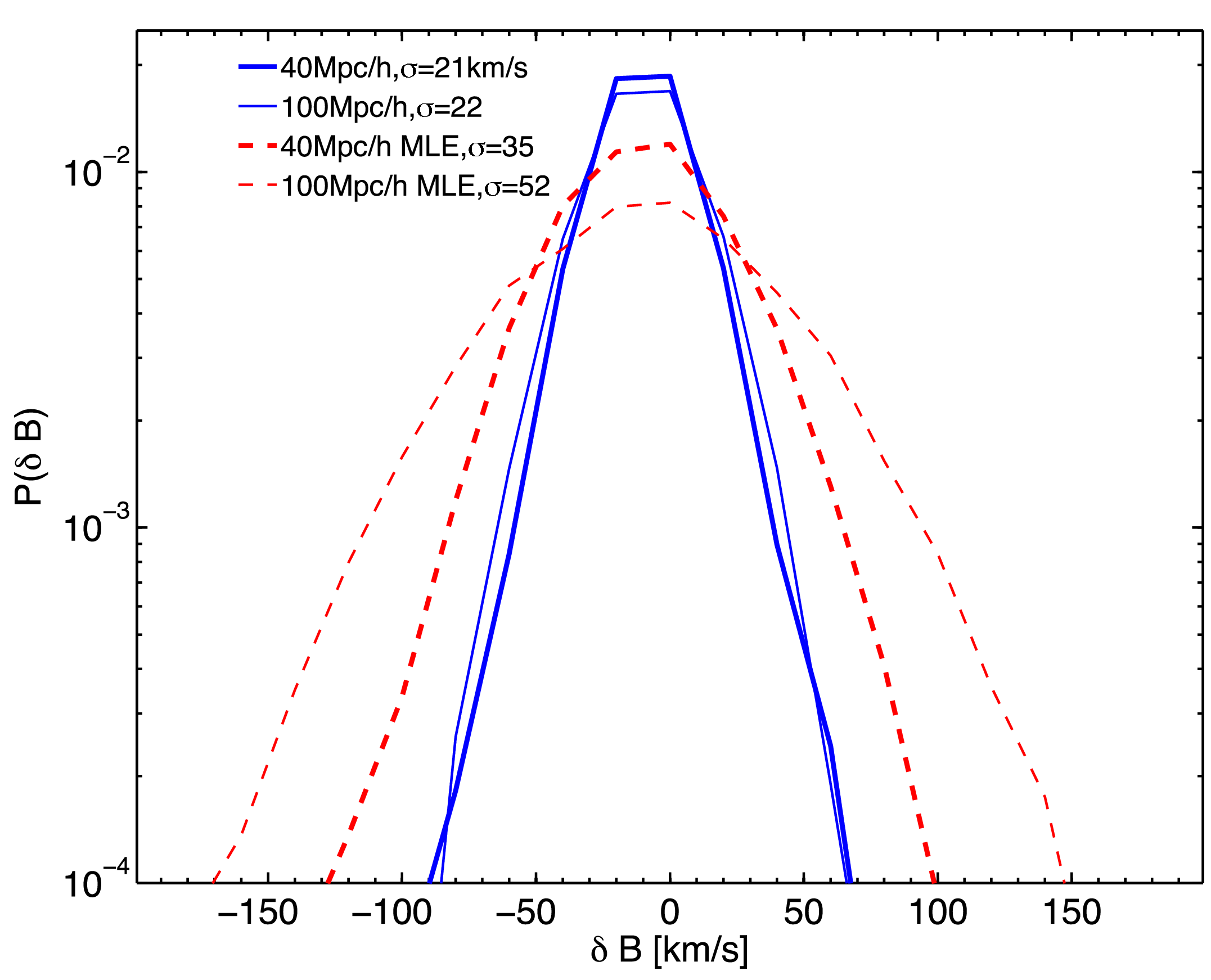}
\caption{The differential distribution functions of the difference between 
estimated and true respective  cartesian components. The notation of the lines
is the same as in the previous figure. }
\label{fig:db}
\end{figure}

In order to test the performance of \ace\ and the MLE reconstructions of the bulk flow 
from the SFI++ TF data we use  2200 mock catalogs of TF measurements. In each of the 
catalogues,  galaxies with the same positions as in the real SFI++ data are assigned absolute magnitudes, $M_i$, and 
line width parameters, $\eta_i$, following 
an artificial ITF relation with slope  $s=-0.12$ and intrinsic scatter  $ \sigma_\eta=0.057$ \citep[e.g.][]{DN10}.
The peculiar velocities of galaxies in each mock are  
taken from a  linear gaussian random velocity field in a cubic box of $1454\hmpc$ on the side.
Each mock is placed randomly in this large box and the peculiar velocity of each galaxy is 
then obtained by interpolating the velocity field on the position of the galaxy. 
A gaussian random realizaton of the velocity field is 
generated for  $\Lambda$CDM7 using the COSMICS package \citep{cosmics}.
Further, we work with  a parametric form of the power spectrum taken  from  \cite{EH98}  (eqs. 29-31 in their paper)

For \ace\ we still need to construct the basis velocity fields $\vv^\alpha(\vx)$. Here we use $N_{\rm a}=120 $ basis 
functions, extracted in a similar way to the mocks, but from a completely different random realisation 
of a velocity field in a very large box.  Thanks to the regularization term in (\ref{eq:chi}), the \ace\ inferred bulk flow has very little dependence on 
the actual value of $N_{\rm a}$ as long as it is large enough to capture the 
main features of the 3D  flow: very similar results are obtained with $N_{\rm a}=50$ and $N_{\rm a}=120$.
Each of these 120 velocity fields $\vv^\alpha(\vx)$ are further smoothed with a gaussian window of 
$10\hmpc$ in width. The purpose of this small scale smoothing is to filter out low frequency modes
which would be over-fitted by the data especially at large distances. This smoothing, however, has very little effect on the 
 bulk flows reconstructed by \ace.
 We emphasize that a  basis function $\vv^\alpha(\vx)$ is not only defined at the 
galaxy positions, but also at any point in a sufficiently large volume (radius $100 \hmpc$) which contains
all the galaxies used in the analysis. Hence, for each basis function we can measure its corresponding 
bulk flow, $\vB^\alpha(r)$, and once coefficients $a^\alpha$ have been determined from the data 
by minimization of (\ref{eq:chi})  then the \ace\ reconstructed  bulk, $\vB_{_{\rm ASCE}}(r)$, is readily given by (\ref{eq:best}).

Figure \ref{fig:scatt} shows   scatter plots of  the estimated versus true bulk flows of a spherical region of 
$60 \hmpc$ in radius. 
 For clarity, results  from only 
400 randomly selected mock catalogues are shown. 
Blue dots and red plus signs correspond to $\vmle$ and $\vace$, respectively.
Because of the anisotropic distribution of the observed galaxies, 
the methods may not reconstruct the three cartesian components equally well. Hence,  $x,y$ and $z$ bulk flow components in Supergalactic coordinates are, respectively, shown in the top, middle and bottom panels. 
Blue and red lines in each panel are linear regressions of the estimated on true bulk flows. The corresponding 
mathematical expressions of the regressions are indicated in each panel.

The regularization term in (\ref{eq:chi}) naturally tends to  underestimate  the coefficients $a^\alpha$ and subsequently the 
reconstructed bulk flow. 
However, the agreement between  the \ace\ reconstructed and the true bulk flows 
seen  in figure (\ref{fig:scatt}) clearly demonstrates that the effect is meagre.
 To further explore the quality of the \ace\ and MLE reconstructions and 
 to ascertain that the regularization term does not cause  a significant reduction in the amplitude of 
 the bulk flow, we  
apply   \ace\ and MLE to the  mock data but with true velocities amplified by a factor of 1.5. Everything else, including the 
  regularization term in (\ref{eq:chi}), remained the
same. The reconstructed $\vace$ and $\vmle$     versus  true amplified bulk flow are  shown in figure (\ref{fig:scatt15}).
 Both \ace\ and MLE perform
well even with this amplification of the bulk flow in the mocks.

In both  \ace\ and MLE, the slopes of the regression lines plotted in  (\ref{fig:scatt}) are close, but not equal to unity. The deviation from unity is significant (compared to the scatter of the points) and persists 
when the regression is done using all  the 2200 mock points. This small but statistically significant bias depends on the radius 
of the sphere for which
bulk flow is computed.  The bias can  easily be calibrated using the mock catalogues. 
Hereafter, all reconstructed  bulk flows, from \ace\ and MLE,  are corrected for the systematic bias in the mean 
of the estimated bulk given the mean of the true value. In practice we 
write the corrected estimate of the bulk flow $\vB_{_{\rm  corr}}(r)$  from the raw bulk $ \vB_{_{ \rm raw}}(r)$ (directly reconstructed by either \ace\ or MLE) as $\vB_{_{\rm corr}}(r)=C_1 \vB_{_{ \rm raw}}(r) +C_2$ where $C_1$ is the 
ratio of the rms values of the true to raw bulk flows and $C_2$ is a constant term which accounts  for the offset between 
the true and raw bulk flows.

In all panels of figure (\ref{fig:scatt}),  the mock  $\vace$ are tightly scattered around their 
corresponding regression lines. The scatter in $\vmle$ appears to be more 
significant. 
 To quantify the scatter between the reconstructed and true bulk flows, 
we plot, in figure (\ref{fig:costh}), the cumulative fraction,  $P<(\theta)$,  of mock catalogues for which the angle between 
estimated and true bulk flows is less than $\theta$. The solid blue and red dashed curves refer to  \ace\ and  MLE, respectively, while thick 
and thin to bulk flows within $40\hmpc$ and $100\hmpc$, also respectively.
The curves are computed after employing the correction to the systematic bias as explained above. 
The performance of \ace\ is excellent.  For $40\hmpc$, the direction of $\vace$ 
is recovered within $3^\circ$ for about $68\%$ of the mocks. For $100\hmpc$ this uncertainty increased to $7^\circ$. 
The \ace\ method is significantly  superior to MLE. The thin blue and thick dashed lines almost overlap, meaning that the performance of 
\ace\  for $100\hmpc$ is as as good as that of MLE for $40\hmpc$. For $r=100\hmpc$, MLE recovers the direction 
only within $27^\circ$ for $68\% $ of the mocks. 

Figure (\ref{fig:db}) plots the differential probability distribution function, $P(\delta B)$, where $\delta B$ 
refers to the difference in all cartesian components between estimated and 
true bulk flows, from the 2200 mocks.  The notation of the lines is the same as in the previous figure and 
as displayed in the figure.  The figure also indicates $\sigma$, the rms value of $\delta B$ for the
plotted cases.  The low values of $\sigma$ corresponding to 
\ace\ demonstrate its excellent ability at recovering the 
true bulk. The performance of MLE is good, but   less satisfactory. 
\section{Results}
\label{sec:res}
\begin{figure} 
\centering
\includegraphics[ scale=0.45 ,angle=00]{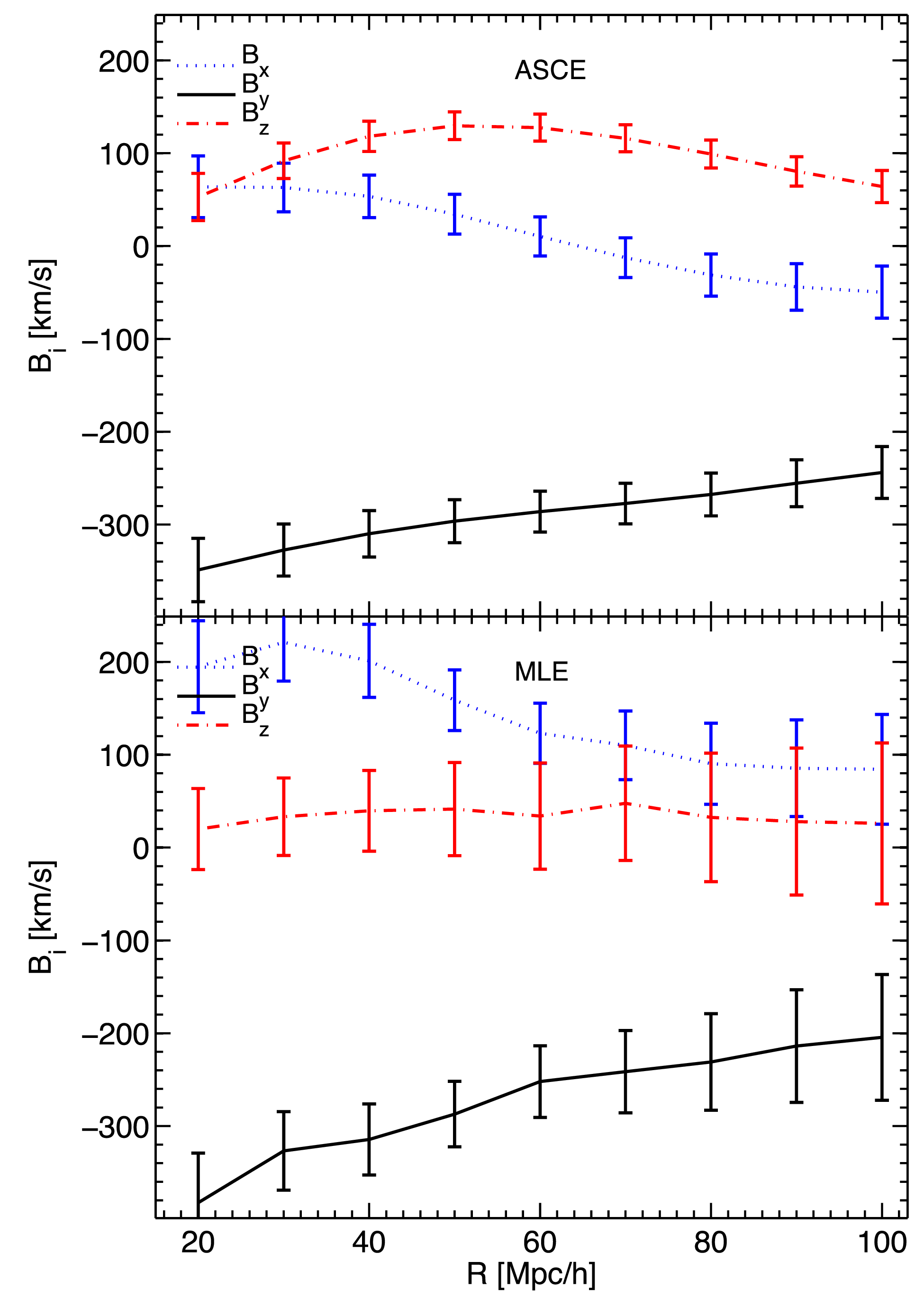}
\caption{ The three Galactic cartesian components of the bulk flow  as a function of radius.
{\bf Top} and {\bf bottom} panels correspond to ASCE and MLE estimation, respectively.  }
\label{fig:unb}
\end{figure}

\begin{figure} 
\centering
\includegraphics[ scale=0.5 ,angle=00]{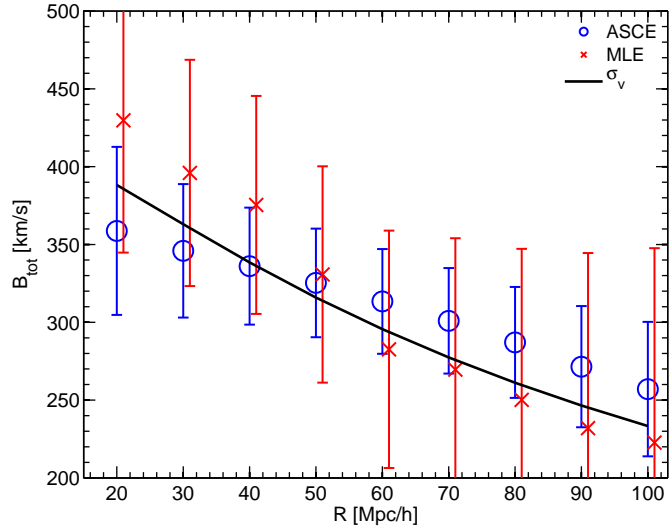}
\caption{ The amplitude of the bulk as a function of distance for ASCE and MLE as indicated in the figure.
The solid curve shows the rms value of the bulk flow as expected in a flat Universe $\Lambda$CDM model 
with $\Omega_m=0.266$, $h=0.71$ and $\sigma_8=0.85$.}
\label{fig:btot}
\end{figure}

The results of the application of the \ace\ and MLE methods to recover the bulk flow
 from 
the real  SFI++ TF catalogue are summarized in figures (\ref{fig:unb})--(\ref{fig:btot}). 
The bulk flows are reconstructed for spheres centered on the Milky-Way and 
 of radii 
  from $r=20\hmpc$ to $100\hmpc$ in steps of 
$10\hmpc$. The smallest radius is chosen large enough so that nonlinear effects are not expected to be important \citep{n91}, 
facilitating the comparison with cosmological models. The largest radius corresponds to  
the distance within which the data are used.
Figure (\ref{fig:unb})  shows the Galactic $x$ (blue dotted), $y$ (black solid), and $z$ (red dot-dashed) components of $\vace$ (top panel) and $\vmle$ (bottom) 
as a function of $r$.  
The magnitudes of $\vace$ and $\vmle$ versus  radius are plotted, respectively,  as the blue circles and  plotted in  figure
(\ref{fig:btot}). 
The $1\sigma$ errorbars in both figures are based on the 2200 mocks. 
  The component $B_y$ is clearly the most significant. This is just a coincident and  has no bearing on the 
statistical analysis of the results as one can always choose a coordinate system such that the bulk is 
along a given axis.The solid  curve  in this figure is the theoretical expectation of  $\Lambda$CDM7 with 
but with $\sigma_8=0.85$ instead of the default $\sigma_8=0.8$. The theoretical curve 
is computed given the density power spectrum $p_{_\delta} (k,\Omega_m,\Omega_b,h,n_s)$ by 
\begin{equation}
\label{eq:sigv}
\sigma_v^2(r)=\frac{H_0^2 f^2}{2\pi^2}\int \dd k \; p_{_\delta} (k) W^2(k r)
\end{equation}
where $f =\Omega_m^\gamma$   with a growth index  $\gamma\approx 0.55$ for a flat Universe \citep{Lind05}, and
 $W=W_{\rm TH}(k r){\rm exp}(-k^2R_s^2/2)$ with $W_{\rm TH}$ is the top-hat window function
and the gaussian window  takes care of 
the fact that the basis functions $\vv^\alpha$ used in \ace\ are smoothed with a gaussian window  of $R_s=10\hmpc$ in width.  
The expression (\ref{eq:sigv}) is obtained assuming the linear relation
$H_0 f \delta =-{\bf \nabla} \cdot \vv(\vx)$  between the density contrast, $\delta$, and $\vv(\vx)$ \cite{Peeb80}.

\begin{figure} 
\centering
\includegraphics[ scale=0.5 ,angle=00]{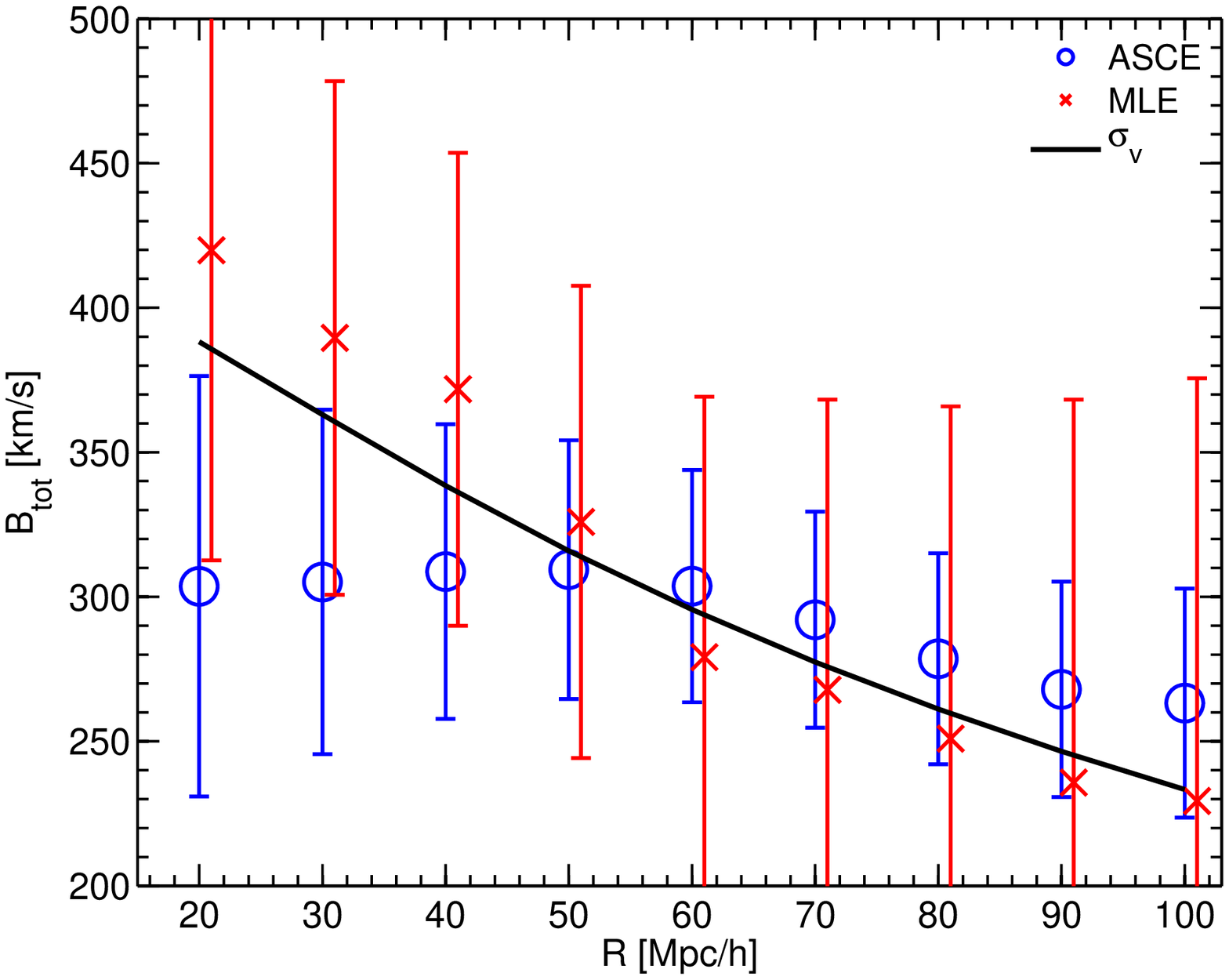}
\caption{ The same as the previous figure with \ace\ basis functions generated using a $\Lambda$CDM7 power spectrum but 
with scalar index $n=0.75$ and $\sigma_8=1$, which has more power on large scales compared to our standard choice $n=0.963$. }
\label{fig:btotn}
\end{figure}

The MLE and \ace reconstructed bulks are similar especially at large distances $r>40\hmpc$. This is because, the data covers 
space more isotropically at larger distances. 
Figure (\ref{fig:scatt15}) clearly demonstrates that \ace\ will not  cause a significant artificial 
under-estimation of  large bulk flows such as reported in \cite{feldwh10}. 
To further, ascertain  that our \ace\  derived bulk flow  is  robust,  the blue circles in figure (\ref{fig:btotn}) the 
amplitude of the \ace\ bulk flow reconstructed using basis functions generated from  a $\Lambda$CDM7 power spectrum but with a scalar 
index $n=0.75$ and $\sigma_8=1$. The results are very similar to the \ace\ bulk flow shown in figure (\ref{fig:btot}) despite the 
significantly enhanced large scale power.  The agreement is particularly striking at large distances.

\subsection{Comparison  with cosmological models}
 \label{sec:comp}
 Figure (\ref{fig:btot}) indicates that the estimated bulk flows are consistent with the theoretical expectations. 
 But the errors are strongly correlated and a proper statistical analysis must take into account the 
 covariance of the errors. Our large number (2200) of mock catalogues allows a robust 
 determination of  the error covariance function between the bulk flow estimates at different radii.
 Since \ace\ is significantly superior to MLE, we restrict the comparison with models to \ace\ reconstruction. 
 We use all components of $\vace$ estimated at 8 values of $r$ ranging from   $r=30\hmpc$ to $100\hmpc$ in steps of 
 $10\hmpc$. The reason for not considering smaller radii is that the bulk is most robustly 
 constrained independent of the assumed basis functions at $r>30\hmpc$. 
We denote the set of  \ace\ reconstructed cartesian components at these 8 values of $r$ by $\cBt$ and the corresponding 
underlying true quantities by ${\cal B}_t$. We write the probability for observing the set $\cBo$ as 
 \begin{equation}
 \label{eq:pr}
 P( \cBo)=    \int \dd \cBt P(\cBo|\cBt) P(\cBt) \; 
 \end{equation}
 where the probability $P(\cBt) $ for the underlying $\cBt$ is computed within the framework of a cosmological model. Here, we adopt the 
 $\Lambda$CDM model. For gaussian 
 velocity fields, the calculation of $P(\cBt)$ is easily done by integrating standard analytic expressions involving the power spectrum.
 We assume that the probability $P(\cBo|\cBt)$ for $\cBo$ given $\cBt$ is gaussian with error covariance matrix computed from the 2200 
 mocks.  Under these assumptions, the   expression (\ref{eq:pr}) yields 
 \begin{equation}
 \label{eq:pbo}
 P( \cBo)=\frac{1}{\sqrt{(2\pi)^d |\Sigma|}}{\rm exp}\left(-\frac{1}{2}\cBo^T \Sigma^{-1} \cBo\right)\; ,
 \end{equation}
 where $d$ is the number of elements in $\cBo$, i.e. $d=24=8\times 3$; for 8 values of $r$ and 3 cartesian components.
 The $d\times d$ covariance matrix $\Sigma=\Sigma_{\rm o}+\Sigma_{\rm t}$,  where $\Sigma_{\rm o} $ is the covariance of the errors 
 on $\cBo$ and $\Sigma_{\rm t}$ describes  the covariance   of the underlying quantities $\cBt$. The dependence on the cosmological models comes 
 through $\Sigma_{\rm t}$.
 \subsection{Consistency with $\Lambda$CDM7}
We begin by assessing how well  $\Lambda$CDM7  
is consistent with the data. To do that we generated $10^7$ sets, ${\cal B}_{\rm rnd}$,  each containing  $d=18 $ numbers
 selected 
at random from a gaussian distribution given by (\ref{eq:pbo})  computed with $\Sigma_{\rm t}$ for 
$\Lambda$CDM7.
For each of those $10^7$ sets of ${\cal B}_{\rm rnd}$ we compute the corresponding $P({\cal B}_{\rm rnd})$ using (\ref{eq:pbo})
and tabulate the negative of the log of the  probability,  $nlP_{\rm rnd}=-\ln P({\cal B}_{\rm rnd})$. We also compute $nlP_{\rm o}=-\ln P(\cBo)$ for the observed 
$\cBo$ also using $\Lambda$CDM7.
We find that only  26\% of the  $10^7$ values of $nlP_{\rm rnd}$ exceed $nlP_{\rm o}$. Therefore, the $\Lambda$CDM7  cannot be rejected by the bulk flow results. 

\subsection{Independent constraints on $\sigma_8$ and $\gamma$}
The $\Lambda$CDM  expected  amplitude of the bulk flow depends separately on the 
cosmological parameters (see equation \ref{eq:sigv} and  the parametric form for the power spectrum in \cite{EH98}).
But the most significant dependence is on $\sigma_8$
and $\Omega_m$ and hence we restrict ourselves here to deriving constraints on these two parameters only. 
 We compute 
$nlP_{\rm o}$ for a grid of values of $\Omega_m$ and $\sigma_8$ used in $\Sigma_{\rm t}$, maintaining  all other
parameters at their default $\Lambda$CDM7 values.  

\begin{figure} 
\centering
\includegraphics[ scale=0.5 ,angle=00]{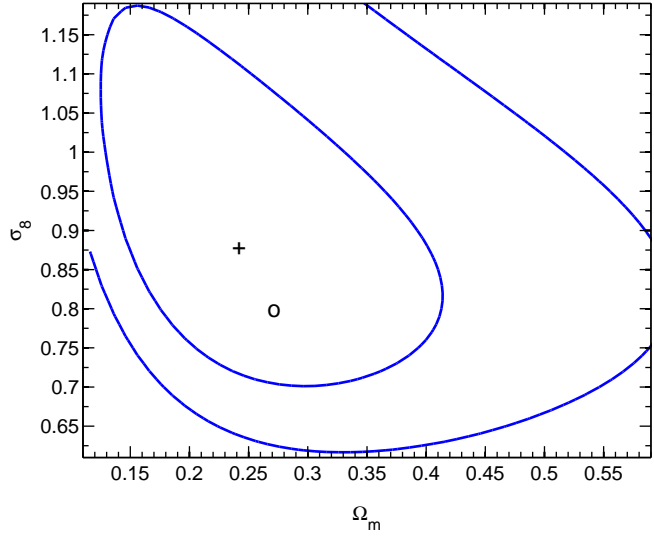}
\caption{ Contour plot of the $68\% $ and $95\%$ confidence levels in the $\Omega_m - \sigma_8$ place. The 
plus sign marks the maximum of the probability distribution function at  $(\Omega_m,\sigma_8=(0.236,0.88)$, while 
the circle indicates  $(0.266, 0.8)$, corresponding to the best fit WMAP7 values.}
\label{fig:sigom}
\end{figure}

\begin{figure*} 
\centering
\includegraphics[ scale=0.5 ,angle=00]{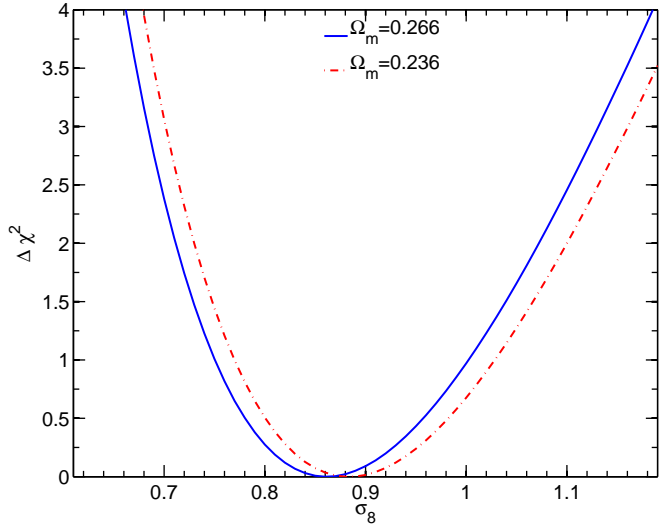}

\caption{  Curves of $\Delta \chi^2$ as a function of $\sigma_8$ for $\Omega_m=0.266$ (blue solid line)
and   $\Omega_m=0.236$ (red dot-dashed).}
\label{fig:sig}
\end{figure*}

\begin{figure} 
\centering
\includegraphics[ scale=0.5 ,angle=00]{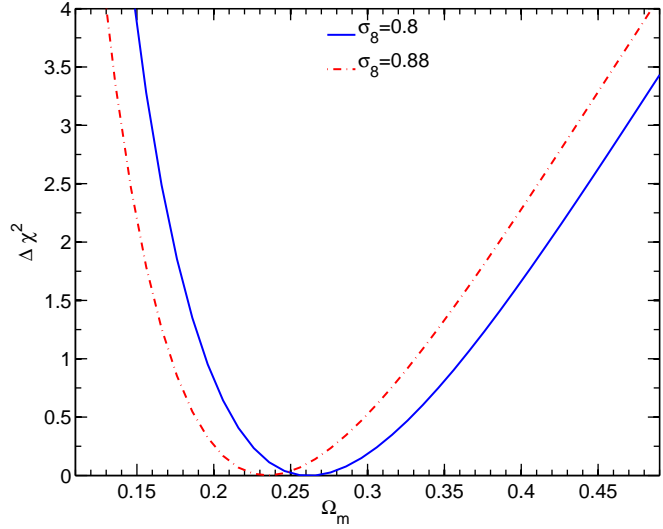}
\caption{  Curves of  $\Delta \chi^2$ as a function of $\Omega_m$ for  $\sigma_8=0.8$ (blue solid) and $\sigma_8=0.88$ (red dot dashed).}
\label{fig:om}
\end{figure}

\begin{figure} 
\centering
\includegraphics[ scale=0.5 ,angle=00]{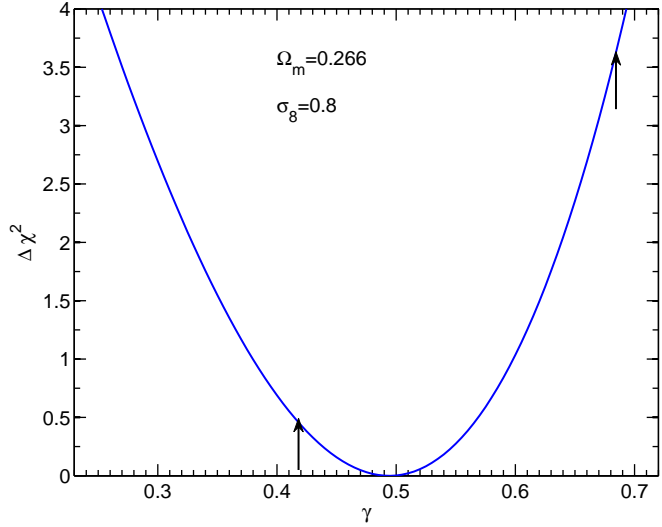}
\caption{  Curves of  $\Delta \chi^2$ as a function of the growth index $\gamma$ given  $\sigma_8=0.8$ and $\Omega_m=0.266$. The left  and right arrows, respectively,  indicate  $\gamma$ values obtained in 
$f(R)$  and flat  DGP models. }
\label{fig:gamma}
\end{figure}

Confidence levels (CLs) on $\Omega_m$ and $\sigma_8$ are obtained by 
inspecting the contours of  $\Delta \chi^2(\Omega_m,\sigma_8) =2(nlP_{\rm o}-{\rm min}(nlP_{\rm o}))$ in the $(\Omega_m,\sigma_8)$ 
plane. The minimum of $ nlP_{\rm o}$ (i.e. $\Delta \chi^2=0$ ) is at 
$(\Omega_m,\sigma_8)=(0.236,0.88)$, marked by the plus sign in the figure.  
 The 
$\Lambda$CDM7 default values $(\Omega_m,\sigma_8)=(0.266,0.8)$  are indicated by the circle. 
The inner and outer contours of   $\Delta \chi^2$ shown  in figure (\ref{fig:sigom})  correspond to 
68\% and 95\% CLs for two degrees or freedom \citep{press}.   The $\Lambda$CDM7 point is  well within the 68\% confidence level.
The shape of the contours implies  
the correlation $\sigma_8\sim \Omega_m^{-0.28}$.  This  reflects  the 
dependence of the shape of the density power spectrum $p_{_\delta}$  on $\Omega_m$ and from the 
factor $f(\Omega) \approx \Omega^{0.55}$ (see eq. \ref{eq:sigv}).  Only if we neglect the dependence of the 
shape of $p_{_\delta}$ on $\Omega_m$ we get  $\sigma_8\sim \Omega_m^{0.55}$.
It is of interest to inspect the constraints when either of the parameters $\Omega_m$ or $\sigma_8$ is fixed at certain 
values. 
Figure (\ref{fig:sig}) shows two curves of $\Delta \chi^2$ versus $\sigma_8$ corresponding  to the WMAP7 $\Omega_m=0.266$ and $0.236$ giving a minimum of $\Delta \chi^2$ in the $(\Omega_m, \sigma_8)$ plane as seen in figure (\ref{fig:sigom}). 
Figure (\ref{fig:om}) plots $\Delta \chi^2$ as a function of $\Omega_m$, for  $\sigma_8$ at the WMAP7 value of  $0.8$ and at $0.88$
corresponding to  the minimum of $\Delta \chi^2$ in figure (\ref{fig:sigom}). 
In  each of the curves in figures  (\ref{fig:sig})  and (\ref{fig:om}), the value of $\Delta \chi^2$ at the minumum of the curve is 
set to zero. Hence, 
the  $\delta \chi^2=1$ and $4$ correspond to 68\% ($1\sigma$) and 95\% ($2\sigma$)  CLs, respectively \citep{press}. 
These curves assume  a growth index $\gamma=0.55$
as is appropriate for a $\Lambda$CDM model. 
Hence figure (\ref{fig:sig}) gives $\sigma_8=0.86\pm 0.11$ ($1\sigma$) for $\Omega_m=0.266$ and $\gamma=0.55$. 
However,  we see from  (\ref{eq:sigv}) that  
by varying $\gamma$ alone we get  the scaling $\sigma_8(\gamma=0.55)=\sigma_8(\gamma)  \Omega_m^{\gamma-0.55}$. We can use this to set a constraint on $\gamma$ if we adopt $\sigma_8=0.8$  and $\Omega_m=0.266$ \citep{wmap7}. Demanding that  $\sigma_8(\gamma)=0.8$, the scaling gives $\gamma=0.496\pm 0.096$. 
Figure (\ref{fig:gamma}) confirms this result. The figure plots $\Delta \chi^2$ as a function of $\gamma$ for 
the adopted values of $\sigma_8$ and $\Omega_m$ as indicated. 
The left and right arrows mark the values $\gamma=0.42$ and 11/16. The lower  value is 
  expected in  $f(R)$ models \citep[e.g.][]{Gann09} and the highest corresponds 
to a Dvali-Gabadadze-Porrati (DGP)  \citep{DGP,Wei} flat braneworld cosmology.
We could also 
 substitute the scaling with $\gamma$ in the correlation  $\sigma_8\sim \Omega_m^{-0.28}$ obtained from 
the contour plot to get the constraint $\sigma_8\Omega_m^{\gamma-0.55}(\Omega_m/0.266)^{0.28}=0.86\pm 0.11$ 
between $\gamma$, $\Omega_m$ and $\sigma_8$.


\begin{figure} 
\centering
\includegraphics[ scale=0.5 ,angle=00]{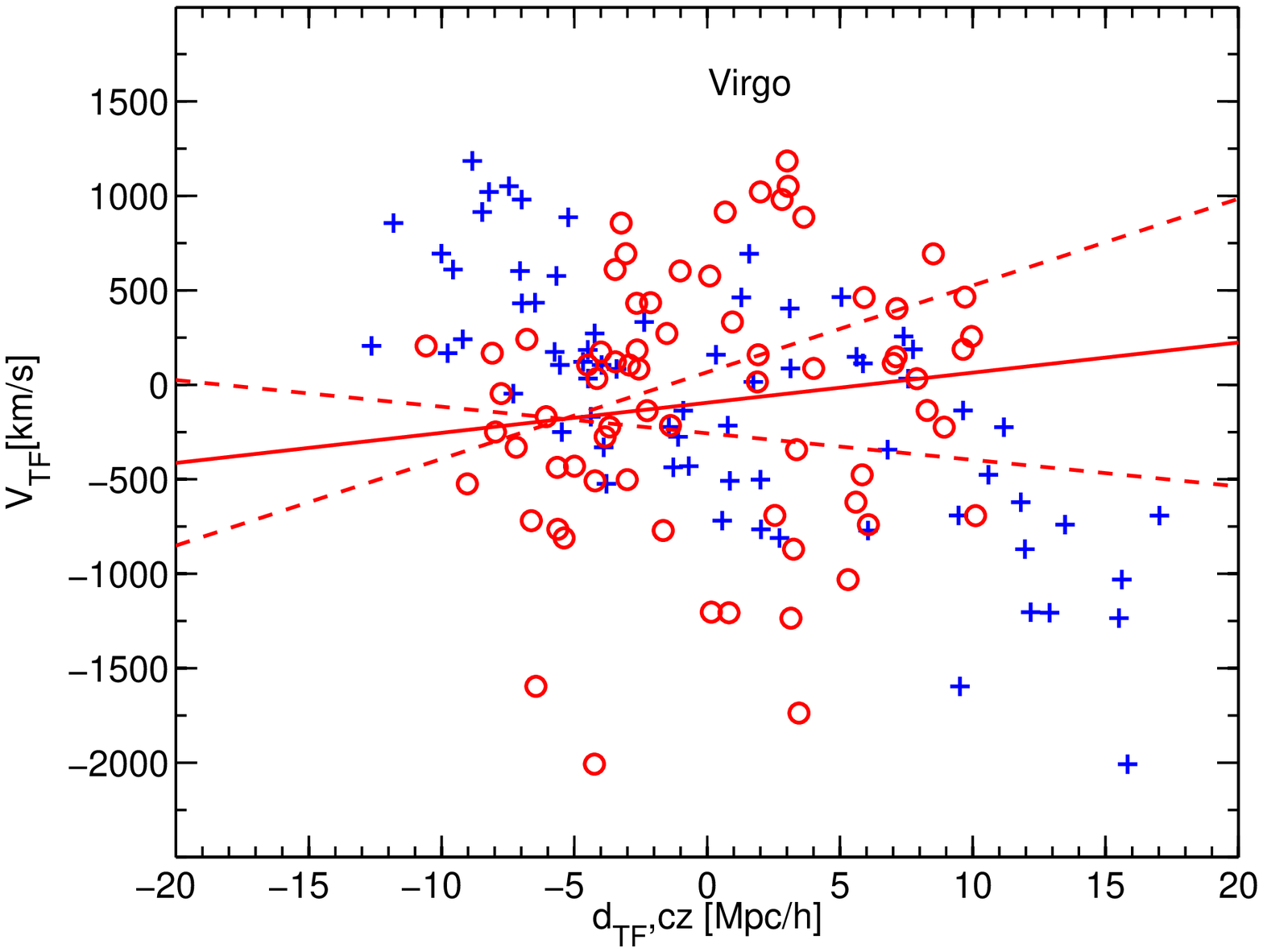}
\caption{Individual peculiar velocities, $V_{\rm TF}$,  of galaxies in the line-of-sight to Virgo, plotted 
against the redshift (red circles) and the estimated distance $d_{\rm TF}$ (blue plus signs).  The centroid of these galaxies is at $cz\sim 0$ and $V_{\rm TF}\sim 0$.}
\label{fig:virgo}
\end{figure}

\section{Discussion}
\label{sec:disc}

The analysis presented here uses a trimmed version of the SFI++ in which galaxies fainter than $M=-20$ are removed. This ensures the linearity of the TF relation. Further, to avoid 
dealing with selection effects imposed on the magnitudes we use the inverse TF relation (see \cite{sw95} for a thorough review of this issue).
Further, to minimize inhomogeneous Malmquist bias \citep{lyn88}, we do not place galaxies at 
their TF inferred  distances, but at their measured redshifts which has significantly 
smaller observational errors.  
We also collapse the main known galaxy clusters. 

The bulk flows estimated here are remarkably featureless and do not seem to reflect 
the gravitational effects of any of the individual main nearby clusters. Bulk flow estimates from 
 TF-like relations are traditionally featureless \cite[e.g.][]{Dek94}, in contrast to velocity dipoles 
estimated by  from the distribution of galaxies in redshift surveys \cite[e.g.][]{nd94}.
In the analysis here, we have collapsed clusters  and therefore,  signatures of individual clusters, in our estimated bulk flows, could be  smeared out. It is instructive to 
explore how much we are missing by collapsing clusters and wether signature of infall on clusters 
of nearby galaxies can actually be clearly seen in  SFI++ or similar data. 
As an illustrative  representative case we plot in figure \ref{fig:virgo} individual 
peculiar velocities of 54 SFI++ galaxies contained in  a cylinder of $6^\circ$ in radius  and of 
$2600\kms$  centered on 
the Virgo cluster. 
 The individual distances, $d_{\rm TF}$  to galaxies are obtained from the observed galaxy deviation from the straight line 
 describing the inverse  TF relation as determined by the 54 galaxies. The 
 individual radial peculiar velocities are then given by $V_{\rm TF}=cz-H_0 d_{\rm TF}$.
Red circles in this figure show  $V_{\rm TF}$ versus $cz$ while the blue plus signs 
correspond to $V_{\rm TF} $ versus $d_{\rm TF}$.
The solid line is obtained by  statistical regression
 of $V_{\rm TF}$ on $cz$, i.e. the red pints. The  two straight dashed lines 
 correspond to 95\% confidence levels of this regression. 
 The  blue points show a pattern that could mistakenly be confused with actual  galaxy  infall onto Virgo. However, this pattern  is entirely due to inhomogeneous 
Malmquist bias: Galaxies scattered to large estimated $d_{\rm TF}$ beyond the cluster, will also have a negative inferred $V_{\rm TF}$.  The effect of this Malmquist bias will be more pronounced for 
more distant clusters  which have larger  absolute errors on $d_{\rm TF}$. 
The red points do not show a clear infall (on Virgo) of  galaxies in the immediate 
vicinity of Virgo. 
 \cite{Kara10} presents an impressive study    of  the observed flow of 1792 galaxies near Virgo.
Taken into account the inhomogeneous Malmquist bias, it is hard to detect a clear infall signature
in the near vicinity of Virgo in this study as all (although the focus of their paper  is different).

The constraints given in figures (\ref{fig:sigom})--(\ref{fig:gamma}) show that the bulk flow alone 
provides useful additional constraints on the cosmological parameters. To achieve tighter  constraints one must 
investigate the full information in the peculiar velocity measurements.  
This could be done by analysis of power spectra  and correlation functions 
  by maximum likelihood techniques \citep[e.g.][]{gd89, jk95, zaroubi97,jusf00,brid01,abate09}.    However, the bulk flow is particularly  appealing because of its simplicity and the fact that it is entirely linear for 
sufficiently large spheres. 
The constraints from peculiar velocities, including the bulk flow, are unique since they are local at redshifts very close to zero and they directly probe the growth
 index $\gamma =\frac{\dd \ln f}{\dd \Omega_m}$ where $f$ is the linear growth factor \citep{Peeb80,Lind05}.
Adopting the WMAP7 cosmological  parameters \citep{wmap7}, we derive a
 local constraint $\gamma=0.495\pm 0.096$.
This constraint is  completely  independent of the biasing relation between 
galaxies and mass. Further, it is essentially a constraint at $z=0$. In contrast, the lowest redshift constraint obtained from a study 
of redshift distortions in the 2dF galaxy redshift survey  is
at $z\approx 0.15$ \citep{Hawk03}. Our constraint significantly improves on previous constraints on $\gamma$ 
\citep{Dos10,Wei} derived at 
 higher redshifts.
This result could help us distinguish between alternative theories for structure formation 
\citep[e.g.][]{amend05,Guz08,knp10}.
For $\sigma_8=0.8$, the constraint on $\gamma$  disfavors DGP models at $\sim 2\sigma $ level, but  it is consistent with  $f(R)$ gravity models \citep[e.g.][]{Starob07,Gann09,Wu09,Fu10}. 
But $\sigma_8$ in these models should be computed self-consistently, assuming the same normalization at the recombination epoch. 
Based on WMAP7, this implies $\sigma_8=0.63$
and $0.855$ for DGP and $f(R)$ models, respectively. Adopting these  $\sigma_8$ values for these models we get 
$\gamma=0.315\pm 0.091$ and  $0.55\pm 0.098$, respectively, for SGP and $f(r)$ (Cinzia Di Porto, priv. comm). 
In the DGP model, the expected  value for  $\gamma$ at $z=0$ is  0.664 \citep{Wu09}, which is ruled 
by our constrain at more than the 3 $\sigma$ level. The $f(R)$ model cannot be ruled out at high confidence level 
by the constraint derived here.

Our results are in agreement with the analysis of \cite{st10}. Peculiar velocities from supernovae, although very sparse, yield bulk flows that are  consistent with WMAP7 \citep{Dai11,colin11}, as we do. 
The results are in agreement with the WMAP7.  The analysis of \cite{Bil11}  of the 2MASS dipole 
from galaxy fluxes is also in agreement with the WMAP7 LCDM. 
The direction  of the bulk flow is robust and agrees with the direction of the motion of the  local group (LG) after 
correcting for the Virgocentric infall  \citep{st10}. But we disagree strongly with the bulk flows \cite{feldwh10}
who find a significant large bulk flow at $r=100\hmpc$, using the untrimmed SFI++ survey, other individual data sets and  also using a composite catalogue. They also use the TF estimated distances instead of the redshifts in their 
analysis of the bulk flow, leading to results which are  highly susceptible to Malmquist bias. 
We have opted to use a single 
uniformly calibrated catalog, namely the SFI++, excluding faint galaxies which spoil the linearity of the ITF \citep{DN10}.
We have also refrained from using composite data since 
 minor  mis-calibration errors between different 
catalogs could  lead to large artificial flows when these catalogues are combined. 
Further, we place galaxies at their measured redshifts rather than estimated distances from the TF relation. This greatly suppresses inhomogeneous Malmquist bias which is known to lead to significant 
spurious signal especial at large distances.
We also refrained from using gaussian window so that the bulk flow within a certain radius
is completely unaffected by the increasing uncertainties at large distances.

We have seen that the MLE and the \ace\ methods give very similar results. Further,
the \ace\  bulk flow at $r>30\hmpc$ is almost completely independent of the cosmological  model 
used in generating the basis functions. This is clearly demonstrated by the comparison of figures
\ref{fig:btot} and \ref{fig:btotn}.
However, in \ace\ even if the results turned out to be sensitive to the assumed model used in generating the basis function, the validity of the model can still be confidently assessed. 
The reason is that the sensitivity would  imply that the data are insufficient for constraining the bulk flow within the framework of the assumed model used in generating the basis function.
Fortunately, this ambiguity is irrelevant for the SFI++ used here since the corresponding bulk flow 
is extremely insensitive to the model used in generating the basis functions.

\section{Acknowledgments}
Special  thanks are due to  Enzo Branchini for many stimulating discussions. 
We thank Cinzia Di Porto for providing $\sigma_8$ for the DGP and $f(R)$ models.
 This work was supported by THE ISRAEL SCIENCE FOUNDATION (grant No.203/09), the German-Israeli Foundation for 
Research and Development,  the Asher Space Research
Institute and  by the  WINNIPEG  RESEARCH FUND.
 MD acknowledges the support provided by the NSF grant  AST-0807630.

\bibliography{bulk_apj}
\appendix 
\section{The regularization term}
The motivation for the second term on the r.h.s in equation  (\ref{eq:chi})  can be found in a 
a  Bayesian statistical formulation.  Our model for the three dimensional 
velocity field defined 
at any point in space $\vx$ is given in terms of an expansion, $\vv^M(\vx)=\sum_\alpha a^\alpha\vv^\alpha(\vx)$,
over a $N_a$ basis functions $\vv^\alpha(\vx)$ each corresponding to 
a realization of a random gaussian field with a cosmologically motivated power spectrum.
Given the data, the  probability $P(\vv^M| {\rm data} ) $ for $\vv^M(\vx)$ is
\begin{equation}
P(\vv^M| {\rm data} )\propto P({\rm data} |\vv^M) P(\vv^M)\; ,
\end{equation}
where 
$P({\rm data} |\vv^M)$ is assumed gaussian with $-\ln P({\rm data} |\vv^M)=\sum_i
 {\left(s M_{0i}+ P^M_i+\eta_0-\eta_i\right)^2}/{\ssigint}$ with $P^M $ related to the radial component of $\vv^M$ by 
 (\ref{eq:utp}). This probability function gives rise to the first term on the r.h.s in equation  (\ref{eq:chi}). 
 The {\it prior} function $P(\vv^M)$ is the probability for the realization of the particular velocity field model $\vv^M$
independent of the data.
 For a gaussian random field
 \begin{equation}
 \label{eq:lnm}
 -\ln P(\vv^M)=\sum_{\vx,{\vx}',J,J'} v^M_J(\vx) \Xi^{-1}  v^M_{J'}({\vx}')
 \end{equation}
where the indices $J$ and $J'$ refer to the three velocity components and $\Xi(\vx,{\vx}',J,J')$ is the velocity correlation 
function.  Substituting the expansion $\vv^M=\sum_{\alpha =1}^{N_a} a^\alpha \vv^\alpha $ yields 
\begin{equation}
 -\ln P(\vv^M)=\sum_{\alpha,\beta}a^\alpha a^\beta \sum_{\vx,{\vx}',J,J'} v^\alpha_J(\vx) \Xi^{-1}  v^\beta_{J'}({\vx}')\; .
 \end{equation}
 Since $\vv^\alpha$ are all space  independent fields, the terms with $\alpha\ne \beta$ will be negligibly small.
 Since all of $\vv^\alpha$ are generated with the same power spectrum, the term 
 $S=\sum_{\vx,{\vx}',J,J'} v^\alpha_J(\vx) \Xi^{-1}  v^\beta_{J'}({\vx}')$ will be     independent of $\alpha$.  Therefore, 
 $ -\ln P(\vv^M)=S \sum_\alpha (a^\alpha)^2$. If the model is also to represent  a random realization of 
 the same power spectrum as each of the basis functions $\vv^\alpha$ then 
 the sum in (\ref{eq:lnm}) must also 
 be equal to $S$. Hence  $P(\vv^M)\propto P(\{a^\alpha\}) \propto \exp(-\sum_\alpha (a^\alpha)^2/2)$ which leads to the
 second term in  (\ref{eq:chi}).

\end{document}

%% file: definitions.tex
\newif\ifAMStwofonts
\AMStwofontstrue

\def\ssigint{\sigma^2_{\eta}}
\def\cBo{{\cal B}_{\rm o}}
\def\cBt{{\cal B}_{\rm t}}

\def\vv{\bf v}
\def\vB{{\bf B}}

\def\gsim{~\rlap{$>$}{\lower 1.0ex\hbox{$\sim$}}}

\def\simpropto{\lower.2ex\hbox{$\; \buildrel \propto \over \sim \;$}}
\def\ltsim{\lower.5ex\hbox{$\; \buildrel < \over \sim \;$}}
\def\gtsim{\lower.5ex\hbox{$\; \buildrel > \over \sim \;$}}
\def\ltsim{\lower.5ex\hbox{$\; \buildrel < \over \sim \;$}}
\def\gtsim{\lower.5ex\hbox{$\; \buildrel > \over \sim \;$}}

\def\pa{\partial}

\def\vmle{\vB_{_{\rm MLE}} }
\def\vace{\vB_{_{\rm ASCE}} }

\def\kms{\mbox{km\,s$^{-1}$}}

\def\dd{\,{\rm d}}

\def\kms{\ {\rm km\,s^{-1}}}
\def\hmpc{\ {\rm h^{-1}Mpc}}

\def\dd{{\rm d}}

\def\ln{{\rm ln}}
\def\pa{\partial}

\def\pmb#1{\setbox0=\hbox{#1}%
\kern-.025em\copy0\kern-\wd0
\kern.05em\copy0\kern-\wd0
\kern-.025em\raise.0433em\box0}

\def\vv{\pmb{$v$}}

\def\vx{\pmb{$x$}}
\def\vr{\pmb{$r$}}

\def\simlt{\lower.5ex\hbox{$\; \buildrel < \over \sim \;$}}
\def\simgt{\lower.5ex\hbox{$\; \buildrel > \over \sim \;$}}

\newcommand{\beq}{\begin{equation}}
\newcommand{\eeq}{\end{equation}}
\def\beqa{\begin{eqnarray}}
\def\eeqa{\end{eqnarray}}
\def\fixit#1{}
\def\hmpc{h^{-1}\,{\rm Mpc}}

\def\dd{{\rm d}}